%% file: main.tex
\documentclass{mgggarticle}
\usepackage[utf8]{inputenc}
\usepackage[english]{babel}
\usepackage{hyperref}
\usepackage[most]{tcolorbox}
\tcbuselibrary{listingsutf8}
\usepackage{graphicx}
\title{Agent Network Protocol Technical White~Paper}
\author{Agent Network Protocol (ANP) Open Source Technology Community\\Gaowei Chang, Eidan Lin, Chengxuan Yuan, Rizhao Cai, Binbin Chen, Xuan Xie, Yin~Zhang}
\date{May 2025}
\version{1.0}
\usepackage{subcaption}
\newtcbox{\codeinline}{on line, boxsep=0pt, left=1pt, right=1pt, top=1pt, bottom=1pt,
colback=gray!20, colframe=gray!80, boxrule=0.5pt, arc=2pt, fontupper=\ttfamily}
\usepackage[T1]{fontenc}
\usepackage{newpxtext} %
\usepackage{newpxmath} %
\usepackage{hyperref}
\usepackage{changepage}
\begin{document}

\begin{titlepage}

\maketitle

\begin{abstract}
With the development of large models and autonomous decision-making AI, agents are rapidly becoming the new entities of the internet, following mobile apps. However, existing internet infrastructure is primarily designed for human interaction, creating data silos, unfriendly interfaces, and high collaboration costs among agents, making it difficult to support the needs for large-scale agent interconnection and collaboration. The internet is undergoing a profound transformation, showing four core trends: agents replacing traditional software, universal agent interconnection, native protocol-based connections, and autonomous agent organization and collaboration.

To align with these trends, Agent Network Protocol (ANP) proposes a new generation of communication protocols for the Agentic Web. ANP adheres to AI-native design, maintains compatibility with existing internet protocols, adopts a modular composable architecture, follows minimalist yet extensible principles, and enables rapid deployment based on existing infrastructure. Through a three-layer protocol system—identity and encrypted communication layer, meta-protocol negotiation layer, and application protocol layer—ANP systematically solves the problems of agent identity authentication, dynamic negotiation, and capability discovery interoperability.

At the application layer, ANP builds an AI-friendly native data network through Agent Description Protocol (ADP) and agent discovery protocol, enabling agents to open capabilities and interconnect in a structured manner. In terms of security and privacy design, ANP introduces human authorization and agent authorization distinction mechanisms, supports multiple DID privacy protection strategies, and promotes minimum information disclosure and end-to-end encrypted communication, ensuring user and agent autonomy and data security in an open network environment.

Looking to the future, ANP is committed to breaking digital silos and driving the internet's evolution from closed platform ecosystems back to a collaborative network centered on open protocols. It upholds the design philosophy that "connection is power," releasing the enormous potential of collective intelligence by promoting high-quality, low-friction agent connections and collaboration. ANP is not just a communication standard but represents an important infrastructure for reshaping openness, fairness, and innovation vitality in the era of the Agentic Web. We invite developers, researchers, and organizations worldwide to participate in building the ANP ecosystem and co-create a new internet future driven by agents.
\end{abstract}

\newpage

\newpage
\tableofcontents

\begin{contributors}
This white paper is a reformatted version of the open-source community edition previously released by the \href{https://github.com/agent-network-protocol}{ANP Open Source Technology Community}. We extend our sincere gratitude to all individuals who contributed their time and expertise (listed alphabetically by last name):
\begin{itemize}
    \item Primary Authors: Gaowei Chang, Eidan Lin, Chengxuan Yuan
    \item Typesetting \& Integrators \& Reviewers: Rizhao Cai, Binbin Chen, Xuan Xie, Yin Zhang
\end{itemize}
\end{contributors}

\end{titlepage}

\input{Sections/section1}

\input{Sections/section2}

\input{Sections/section3}
\input{Sections/section4}
\input{Sections/section5}

\newpage


\bibliographystyle{ieeetr}
\bibliography{ref}
\end{document}

%% file: Sections/section1.tex
\section{Introduction}
\subsection{Background and Challenges}
With the development of large models and autonomous decision-making AI, agents are becoming the new entities of the internet and mobile apps. Agents are expected to replace existing software and become an important part of the internet. However, current internet technology foundations and connection paradigms struggle to meet the needs of agents, mainly in the following aspects:

\begin{itemize}
    \item Data Silos and Context Limitations: Agents need to integrate comprehensive contextual information to make decisions, but current internet platforms are disconnected, with data isolated in different applications and servers. This data silo phenomenon limits agents' access to complete information, making it difficult to leverage their intelligent decision-making advantages.

    \item Unfriendly Interfaces (Non-Native): Existing internet applications are primarily designed for humans, providing services through GUIs. If agents want to use existing human interfaces, they need to simulate human operations, which is inefficient and error-prone. Agents are better at directly using APIs or communication protocols to interact with the digital world, thus requiring native data interfaces for agents to reduce information processing intermediaries and improve interaction efficiency.
    
    \item  High Collaboration Costs: Currently, different agents are isolated from each other, lacking direct communication mechanisms. Although large language models enable agents to negotiate using natural language, without standard protocol support, self-organizing collaboration remains challenging. If any agent could communicate directly with others and autonomously negotiate, it would create a more cost-effective and efficient collaborative network than the existing internet.
\end{itemize}

\subsection{Four Trends of the Agentic Web}
\subsubsection{Agents Will Completely Replace Traditional Software}
In the future, agents will gradually replace existing software applications and become an important infrastructure of the internet. At the individual level, AI assistants will replace the vast majority of existing apps, becoming the main entry point for users to access the internet. Compared to traditional apps, personal AI assistants can achieve order-of-magnitude improvements in information integration, decision support, and scenario interaction experiences. At the enterprise level, companies will connect and interact directly with users on the internet by deploying agents, providing precise and efficient services. Meanwhile, a new connection paradigm represented by point-to-point, de-platformized connections between personal assistants and enterprise agents is taking shape.

\subsubsection{Universal Interconnection Between Agents}
The second core trend of the Agentic Web is to achieve free connections between any agents, which will completely break the current internet's data silo pattern and enable the free flow of information. A fully connected agent network allows AI to fully access cross-domain, cross-platform complete contextual information, thereby helping individuals and enterprises make more comprehensive and precise decisions. At the same time, this open connection mode allows agents to call upon all tool capabilities across the network, greatly expanding the depth and complexity of agent collaboration, making interactions between agents the most mainstream connection method in the future internet.

\subsubsection{Protocol-Based Native Connection Mode for Agents}
Currently, AI's interaction with the internet is mainly through human-centered design methods such as browsers or software interfaces (Computer Use / Browser Use). However, these methods are only temporary transitional solutions and cannot fully release AI's potential. AI is essentially better at directly processing underlying structured data and semantic information rather than human interfaces and webpage HTML. Therefore, we believe that in the future, agents will connect through communication protocols (Protocol) specifically designed for AI native use, which will be widely applied like HTTP and become an industry standard. Based on this protocol, a new data network specifically designed for AI, more accessible and operable for agents, will also emerge.

\subsubsection{Agents Can Self-Organize and Collaborate}
The fourth core trend of the Agentic Web is that agents have the ability to self-organize and collaborate. Through standard protocol support, agents can use natural language for flexible automatic negotiation, quickly clarify each other's needs, and dynamically form collaborative relationships to complete complex tasks together. This collaboration mode, which does not require fixed structured interfaces, greatly improves network operational efficiency and task response capabilities, and significantly reduces human intervention and communication costs. As a result, a highly flexible, efficient, and low-cost agent network will gradually form.

Given these trends, we urgently need new communication protocols to support the development of the Agentic Web. Agent Network Protocol (ANP) is proposed against this background, aiming to provide agents with an open and unified connection mechanism, solving the three major challenges of interconnection, native interfaces, and efficient collaboration, thereby releasing the full potential of AI agents.

%% file: Sections/section2.tex
\section{Core Design Principles}
Agent Network Protocol (ANP) has been designed from the outset with the needs of the Agentic Web at its core, striving to build an open, flexible, and implementable agent communication standard. Its design follows these core principles below:

\subsection{AI-Native Design}
ANP is not a protocol designed for human-machine interaction (such as HTML or GUI interfaces), but is natively designed for direct communication between AI agents. It emphasizes structured data, semantic expression, and natural language integration, allowing agents to understand, discover, and collaborate without simulating human operations.

\begin{tcolorbox}[colback=gray!10, colframe=gray!50, boxrule=0.5pt, arc=2mm]
Goal: To enable agents to access the internet world most efficiently and naturally, releasing the potential of AI decision-making and action.
\end{tcolorbox}

\subsection{Compatibility and Reuse}
ANP respects and is compatible with widely used internet protocol standards such as OpenAPI, JSON-RPC, and audio-video protocols (like WebRTC). Without reinventing the wheel, ANP encapsulates these existing protocols into a semantic framework that agents can understand and negotiate through lightweight meta-protocols and description standards.

\begin{tcolorbox}[colback=gray!10, colframe=gray!50, boxrule=0.5pt, arc=2mm]
Goal: While ensuring innovation, minimize learning costs and migration thresholds to promote rapid integration of existing ecosystems into the Agentic Web.
\end{tcolorbox}

\subsection{Composability}
ANP adopts a modular design, with core components such as agent identity (DID), agent description (ADP), and agent discovery that can be used independently or freely combined. Developers can flexibly choose to use a single protocol component or fully integrate based on actual needs.

\begin{tcolorbox}[colback=gray!10, colframe=gray!50, boxrule=0.5pt, arc=2mm]
Goal: Support application scenarios of different complexities and stages, adaptable to everything from minimum viable units to large-scale systems as needed.
\end{tcolorbox}

\subsection{Simplicity and Extensibility}
ANP follows minimalist principles (Keep It Simple), striving for a concise and intuitive protocol that lowers the barriers to understanding and implementation. Simultaneously, it maintains ample extension interfaces and flexible Schema mechanisms to support the natural addition of new functionalities as the agent ecosystem evolves.

\begin{tcolorbox}[colback=gray!10, colframe=gray!50, boxrule=0.5pt, arc=2mm]
Goal: To maintain a small and stable core protocol while reserving growth space for diverse application requirements.
\end{tcolorbox}

\subsection{Pragmatic Deployability}
ANP's design fully considers practical deployment feasibility, operating primarily on existing internet infrastructure (such as DNS, HTTPS, Web servers, DID systems) without relying on complex or difficult-to-popularize new technologies or proprietary components. Any developer or enterprise can quickly build agent services supporting ANP.

\begin{tcolorbox}[colback=gray!10, colframe=gray!50, boxrule=0.5pt, arc=2mm]
Goal: To lower initial deployment barriers and promote rapid piloting, application, and adoption of the protocol in the real world.
\end{tcolorbox}

\subsection{Principle of Least Trust}
No participant in the network (agents, nodes, etc.) should be fully trusted by default. All interactions must be authenticated and authorized, granting only the minimum permissions necessary to complete a specific task. This necessitates robust identity verification, access control, and data validation mechanisms within the protocol design to mitigate potential security risks and abuse.

\begin{tcolorbox}[colback=gray!10, colframe=gray!50, boxrule=0.5pt, arc=2mm]
Goal: To establish an agent network that can operate securely even in untrusted environments.
\end{tcolorbox}

%% file: Sections/section3.tex
\section{Core Protocol Design}
\subsection{Protocol Layer Architecture}
We have designed a three-layer protocol architecture, as shown in Figure~\ref{fig:anp-architecture}:

\begin{figure}[h]
    \centering
    \includegraphics[width=0.8\linewidth]{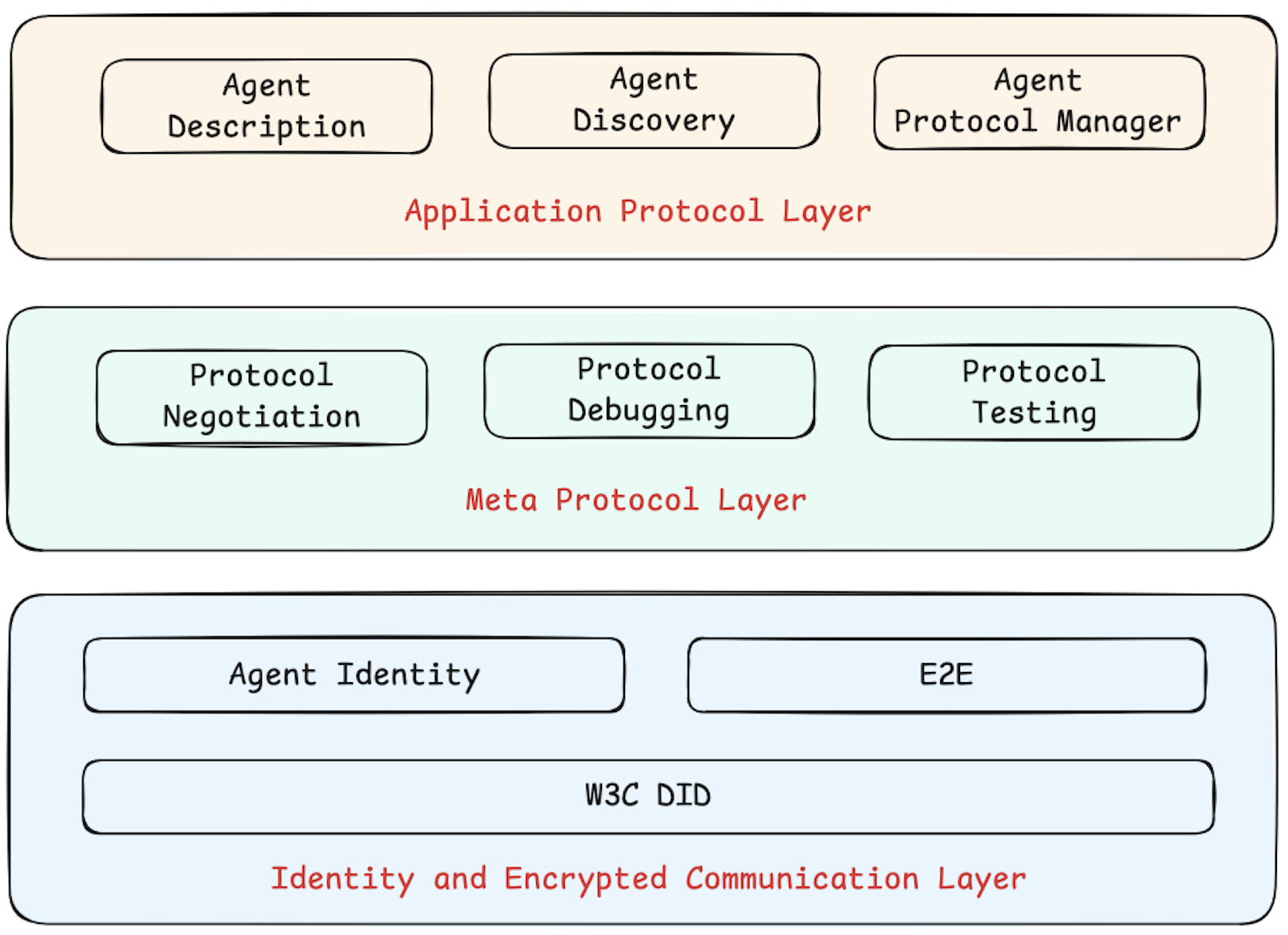}
    \caption{The Architecture of the ANP protocol}
    \label{fig:anp-architecture}
\end{figure}

The ANP protocol architecture is divided into three layers from bottom to top, each with distinct core functions:

\noindent (1). Identity and Secure Communication Layer

This layer defines the basic standards for 
identity authentication and encrypted communication between agents, addressing cross-platform identity mutual recognition and end-to-end encrypted communication issues. ANP, based on the W3C DID standard, has designed a lightweight, extensible decentralized identity authentication mechanism (such as the did:wba method), ensuring that any two agents can securely verify each other's identity and establish private, reliable encrypted communication channels without central authority intervention.

\noindent (2). Meta-Protocol Layer

This layer defines how agents negotiate communication protocols dynamically and adaptively based on natural language. Through natural language exchange of requirements, capabilities, and collaborative intentions, agents can flexibly negotiate communication details (such as request formats, interface calling methods, session management strategies) suitable for the current scenario, achieving highly adaptive protocol coordination and runtime optimization.

\noindent (3). Application Protocol Layer

This layer has two core modules: agent description and agent discovery. Agents publish their functionalities and interfaces through structured description documents (such as the ADP Agent Description Protocol) and expose service entry points actively or passively based on discovery protocols.

\subsection{Identity and Secure Communication Layer}
To achieve interconnection and interoperability among all agents, the primary task is to solve the identity authentication problem between agents. Currently, most internet applications use centralized identity technologies, making cross-system account authentication difficult due to different technical implementations. Although OAuth2.0 technology alleviates this problem to some extent~\cite{ref1-hardt2012oauth}, it was not specifically designed for cross-system identity authentication, making its process relatively complex and limited in decentralization. Therefore, there is an urgent need for a convenient, cross-platform, and decentralized identity authentication technology.

While blockchain-based decentralized identity authentication solutions offer potential approaches, they have not yet become optimal solutions due to scalability challenges in large-scale applications.

\begin{figure}[h]
    \centering
    \begin{subfigure}[b]{0.48\linewidth}
        \centering
        \includegraphics[width=\linewidth]{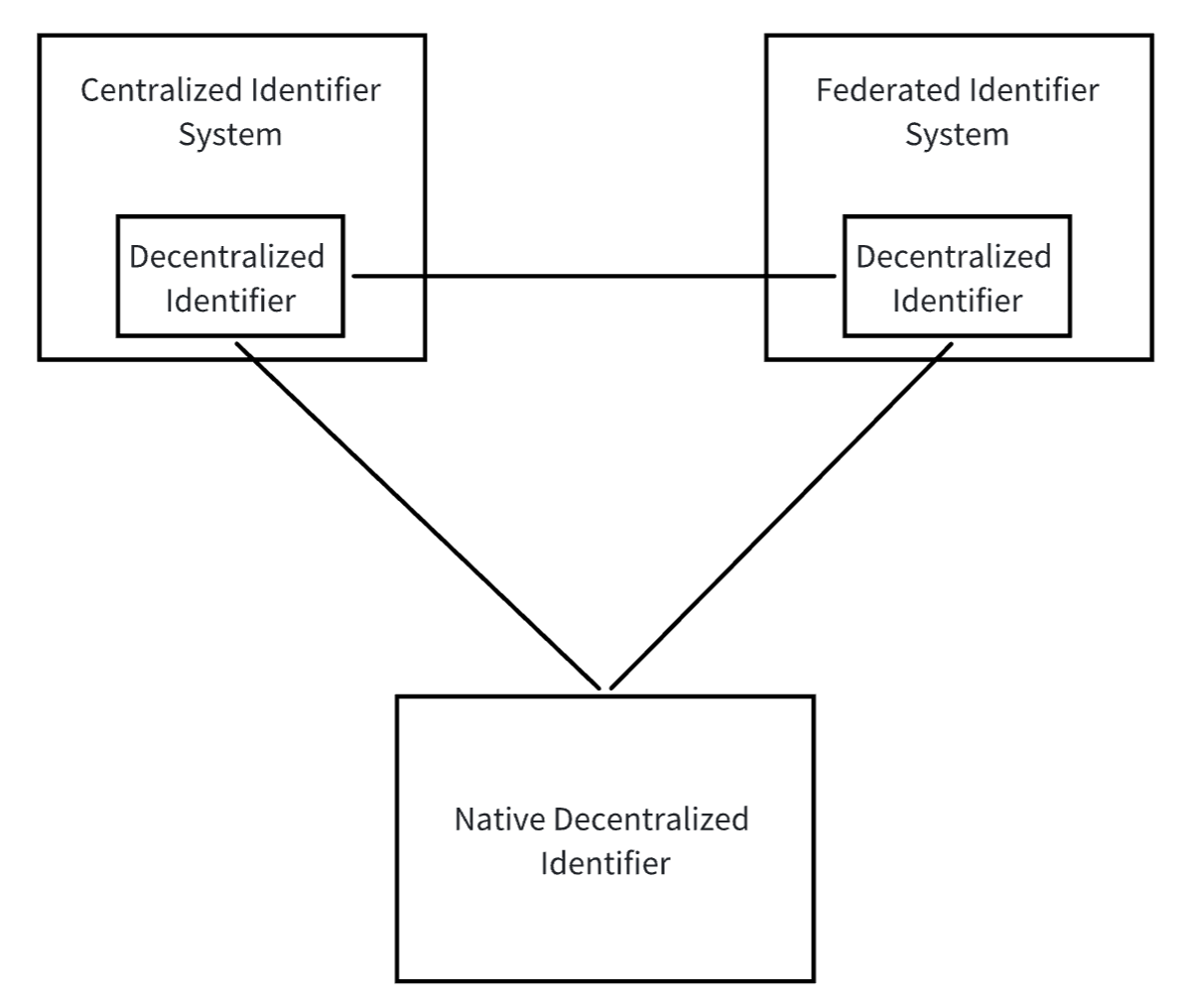}
        \caption{Illustration of DID as identity bridge.}
        \label{fig:did-as-identity-bridge}
    \end{subfigure}
    \hfill
    \begin{subfigure}[b]{0.48\linewidth}
        \centering
        \includegraphics[width=\linewidth]{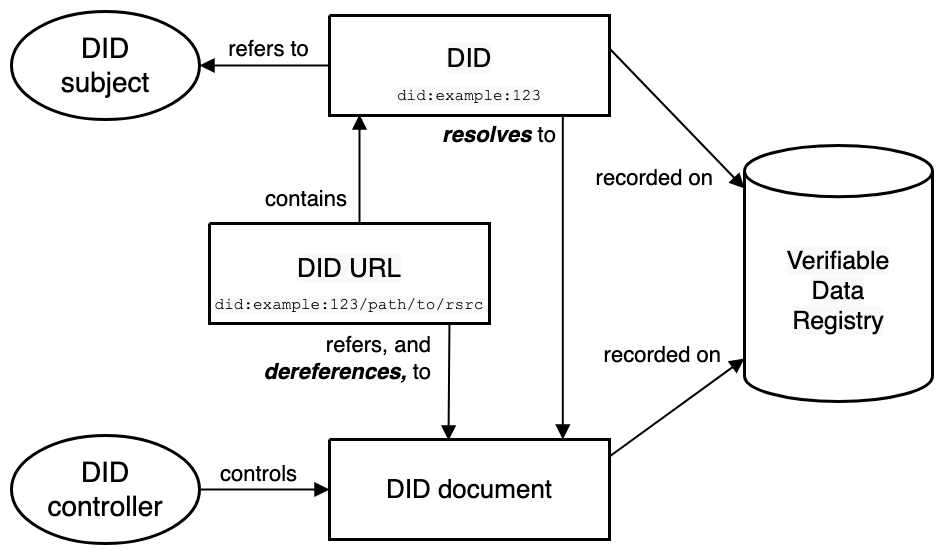}
        \caption{DID diagram.}
        \label{fig:did-architecture}
    \end{subfigure}
    \caption{Overview of DID and its role as an identity bridge.}
    \label{fig:did-overview}
\end{figure}

To address these issues, we introduce the W3C Decentralized Identifier (DID) standard~\cite{ref-2}. As depicted in Figure~\ref{fig:did-overview}, DID is a new identifier standard designed to solve the dependencies of traditional centralized identity management systems. It enables users to control their identities and authenticate each other without relying on centralized systems. The DID core specification does not mandate specific computational infrastructure for building decentralized identifiers, allowing us to fully leverage existing mature technologies and well-established Web infrastructure. Additionally, various types of identifier systems can add DID support, building interoperability bridges between centralized, federated, and decentralized identifier systems. This means existing centralized identifier systems don't need complete restructuring; they only need to create DIDs on their foundation to achieve cross-system interoperability, significantly reducing technical implementation difficulties.

The core component of DID is the DID document, which contains key information related to a specific DID, used to verify the identity of the DID owner and support the management of operations, permissions, and access controls related to the DID.

\begin{figure}
    \centering
    \includegraphics[width=0.8\linewidth]{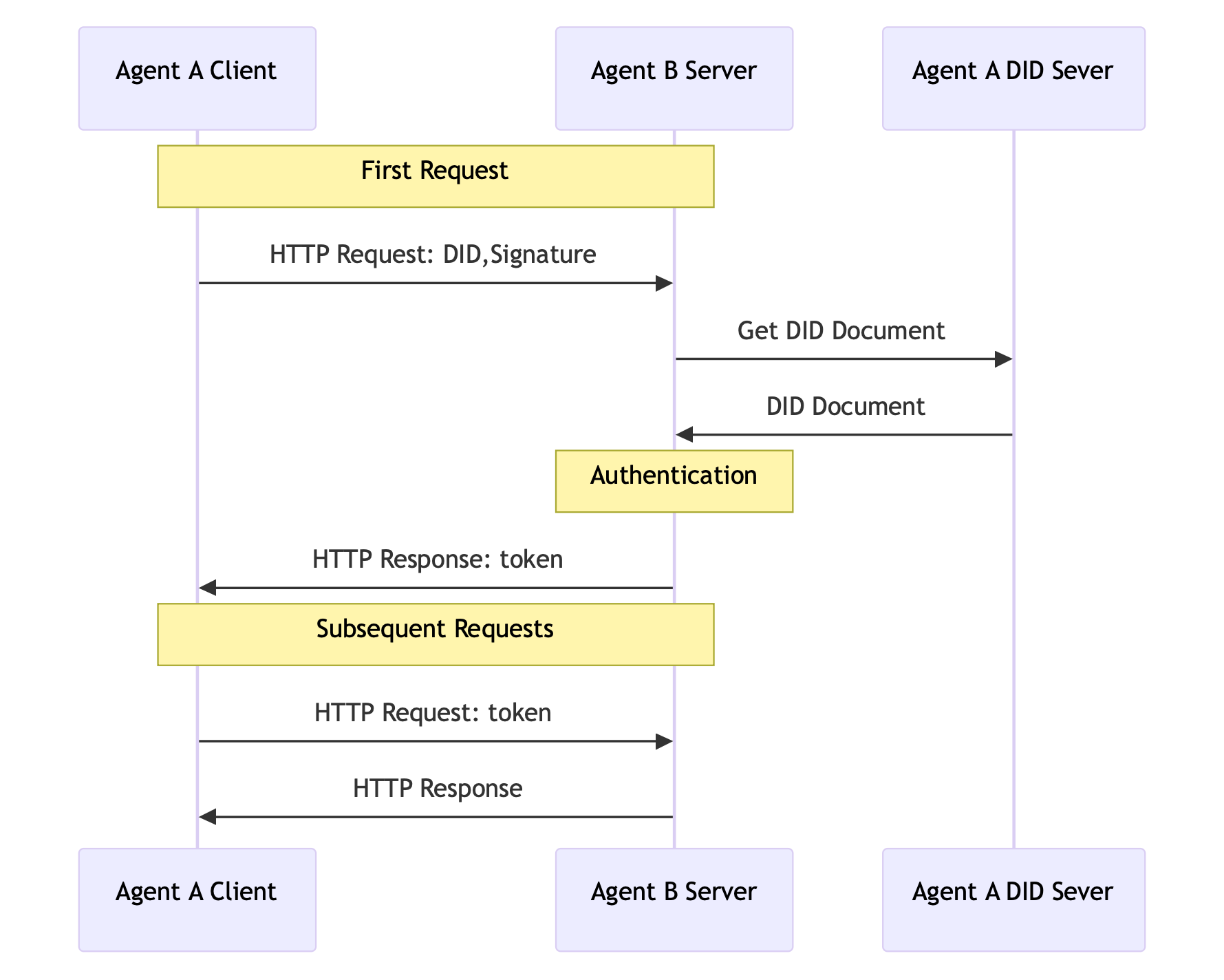}
    \caption{Agent authentication and token exchange sequence.}
    \label{fig:Agent-authentication-and-token-exchange-sequence}
\end{figure}

In the authentication process, as shown in Figure~\ref{fig:Agent-authentication-and-token-exchange-sequence}, the DID document contains methods and corresponding public keys for verifying user identity (private keys are kept by users). Clients can include DID and signatures in HTTP headers during the first HTTP request. Without increasing interaction frequency, servers can quickly verify client identity using public keys from the DID document. After initial verification, the server can return a token, which clients include in subsequent requests, allowing servers to verify the token rather than client identity each time. The core of this process is that verifiers use trusted public keys to verify user signature information, completing identity authentication, permission verification, and data exchange in a single request, making the process concise and efficient.

DID methods define how to create, resolve, update, and deactivate DIDs and DID documents, as well as how to perform authentication and authorization. Among existing DID method drafts, the \codeinline{did:wba} method~\cite{ref-3} is built on mature Web technologies, allowing systems to create, update, and deactivate DIDs and DID documents using centralized technologies (such as cloud computing). Different systems achieve interoperability through the HTTP protocol, similar to email services on the internet, where each platform implements its own account system in a centralized way while enabling interconnection between platforms.

Based on the \codeinline{did:wba} method, we have proposed a new DID method—\codeinline{did:wba}(Web-Based Agent)—specifically designed for agent communication scenarios, adding cross-platform identity authentication processes and agent description services. The \codeinline{did:wba}method inherits the advantages of \codeinline{did:wba} while further optimizing identity authentication mechanisms between agents, enhancing its applicability in agent networks.

Additionally, users typically create one or more public-private key pairs for DIDs, which are used not only for identity verification but also for end-to-end encrypted communication. Based on these DID key pairs, we have designed an end-to-end encryption scheme using the Elliptic Curve Diffie-Hellman Ephemeral (ECDHE) protocol\cite{ref4}, enabling secure communication between two DIDs and ensuring that intermediate nodes cannot decrypt the communication content. End-to-end encryption can be used for agent reverse proxies; for example, an agent may rent ports from third-party platforms to reduce operational costs, where the third-party platform forwards messages but cannot decrypt their contents.

\subsection{Meta-Protocol Layer}
A meta-protocol is a protocol that defines rules for operating, parsing, combining, and interacting with communication protocols. Essentially, it is a protocol for negotiating communication protocols, not directly handling specific data transmission, but providing a flexible, general, and extensible communication framework. Currently, there are two main methods for communication between agents:
\begin{itemize}
    \setlength{\itemindent}{1em} \item Human engineers designing communication protocols: Such as common industry standards. Human engineers design communication protocols for agents, develop protocol code, and perform debugging, testing, and deployment. However, this approach often faces challenges such as high development costs, slow protocol updates, and difficulty adapting to new scenarios.

    \setlength{\itemindent}{1em} \item Agents directly using natural language for communication: Agents communicate using natural language and process this data internally using large language models (LLMs). However, this method has issues with high data processing costs and low processing accuracy.
\end{itemize}

\begin{figure}[h]
    \centering
    \includegraphics[width=0.5\linewidth]{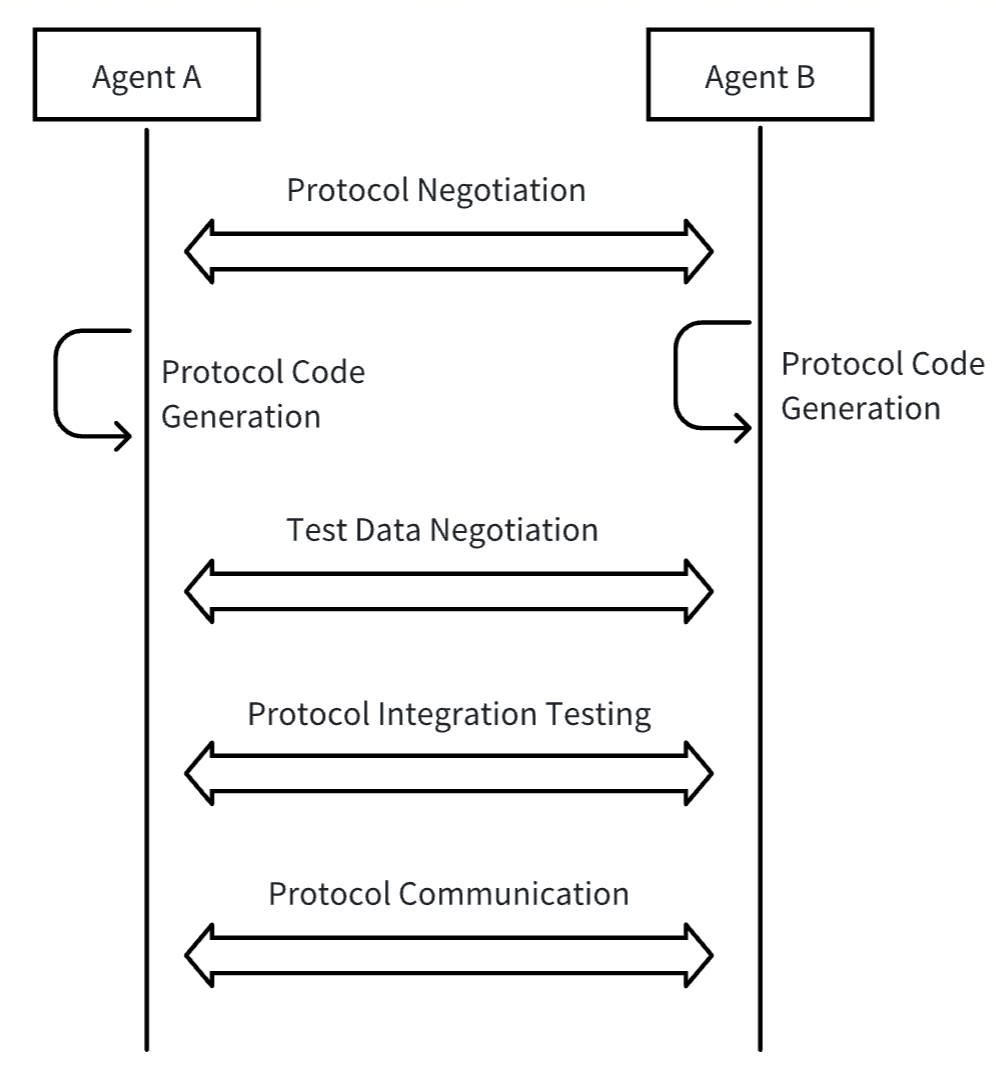}
    \caption{Illustration of Meta Protocol Communication.}
    \label{fig:MPC}
\end{figure}

To address these problems, meta-protocols combined with AI code generation can be used. By using meta-protocols and leveraging AI code generation technology, communication efficiency between agents can be significantly improved, costs reduced, while maintaining flexibility and personalization~\cite{ref5-marro2024scalable}. As illustrated in Figure~\ref{fig:MPC}, the basic process of communication using meta-protocols is as follows:
\begin{itemize}
    \setlength{\itemindent}{1em} \item \textbf{Meta-protocol request}: Agent A first sends a meta-protocol request to Agent B. The request body uses natural language to describe its needs, inputs, expected outputs, and proposes candidate communication protocols. Candidate protocols generally include transport layer protocols, data formats, data processing methods, etc.

    \setlength{\itemindent}{1em} \item \textbf{Protocol negotiation}: After receiving the meta-protocol request, Agent B uses AI to process the natural language description and, combined with its own capabilities, determines whether to accept A's request and candidate protocols. If B's capabilities cannot meet A's request, it directly refuses; if B does not accept A's candidate protocols, it can propose its own, entering the next round of negotiation. This process continues until both parties reach an agreement or negotiation fails.

    \setlength{\itemindent}{1em} \item \textbf{Code generation and deployment}: After reaching an agreement, each party generates and deploys protocol processing code based on the negotiated protocol.
    
    \setlength{\itemindent}{1em} \item \textbf{Joint testing}: After code deployment, both parties negotiate test data to jointly debug and test the protocol and the AI-generated protocol processing code.
    
    \setlength{\itemindent}{1em} \item \textbf{Formal communication}: After testing is complete, the protocol goes live. Thereafter, Agents A and B begin communicating using the finally negotiated protocol and process data using AI-generated code.
    \setlength{\itemindent}{1em} \item \textbf{Handling requirement changes}: If requirements change, the above process is repeated until both parties reach a new agreement.
\end{itemize}

However, meta-protocol negotiation is time-consuming and depends on AI code generation capabilities. If meta-protocol negotiation is performed for each communication, it will result in enormous cost consumption and poor interaction experience. Given that there are many identical or similar communication processes between agents, agents can save the results of meta-protocol negotiations. When similar needs arise later, they can directly use previous negotiation results as formal protocols for communication or as candidate protocols for negotiation. At the same time, agents can also share negotiation results for other agents to query and use.

How to economically incentivize agents to actively upload negotiation results and select consensus protocols between agents is an issue that still requires in-depth research at the meta-protocol layer.

\subsection{Application Protocol Layer}
The application protocol layer has two core modules: agent description and agent discovery.
\subsubsection{Agent Description Protocol (ADP)}
The Agent Description Protocol (ADP) provides a standardized self-description method for each agent. ADP documents are written in JSON-LD (a linked data format based on JSON), using general vocabulary provided by schema.org with extensions, combined with ANP custom vocabulary, to define a unified metadata format for agents. An Agent Description (AD) document is equivalent to the agent's business card or profile in the network world, allowing other agents to understand how to interact with it by reading this document.

\noindent An agent description document typically contains the following key information:
\begin{itemize}
    \setlength{\itemindent}{1em} \item Basic Information: Agent name, unique identifier, affiliated entity (individual, organization, or other agent), and other identifying metadata.
    \setlength{\itemindent}{1em} \item Capability Description: A list of functions, products, or services provided by the agent, along with corresponding descriptions. For example, whether it is a chat assistant, data analysis service, or possesses robot execution capabilities.
    \setlength{\itemindent}{1em} \item Interfaces and Protocols: Methods for interacting with the agent, including callable API endpoints and supported communication protocol versions or specifications. Other agents use this information to know how to send requests or messages to it.
    \setlength{\itemindent}{1em} \item Security and Authorization: The agent's authentication method and permission requirements. For example, which DID identifier and public key are used to verify its identity. The ADP specification recommends using the did:wba method as a unified security mechanism to meet cross-platform identity authentication requirements.
    \setlength{\itemindent}{1em} \item Contact Information: Optional technical support or contact channels (such as service endpoint URLs, maintainer contact emails, etc.) to facilitate human intervention or debugging outside of automatic interactions.
\end{itemize}
The above information is presented in a structured form as a JSON-LD document, making agent descriptions both machine-parsable and semantically rich. Through JSON-LD contexts, vocabularies in AD documents can link to publicly defined ontologies, achieving semantic consistency across different systems. For example, ADP defines core vocabularies such as "Agent," "Service," "Interface," etc., whose meanings are unified for all agents following this specification. This provides a foundation for agents to understand each other's capabilities.

It is worth mentioning that ADP incorporates decentralized identity into the security architecture of description documents. Each agent description can be bound to a DID, with built-in verification methods and public key information for authenticating agent identity. When other agents read the description and communicate with it, they can verify the DID signature to confirm the other party's identity is trustworthy, thereby establishing secure connections in an open network. This design ensures mutual trust between agents, laying the foundation for further collaboration.

Finally, the agent description document serves as an entry point for an agent in the agent network, similar to a website's homepage, allowing other agents to understand how to interact with it by reading the agent description document.

\subsubsection{Agent Discovery Protocol}
With agent descriptions in place, a mechanism is needed for agents to discover each other and establish connections. The Agent Discovery Protocol is designed as a standard protocol for this purpose. It specifies how to publish and retrieve agent description documents in the network, allowing any agent or search service to easily find the entry points of other agents. Essentially, the Agent Discovery Protocol serves as a "search engine protocol" for agents, ensuring that nodes in the agent internet are visible and accessible. The Agent Discovery Protocol provides two complementary discovery methods:
\begin{itemize}
    \setlength{\itemindent}{1em} \item \textbf{Active Discovery}: Based on the common Web well-known path convention, this allows querying all publicly available agent descriptions under a known domain. Specifically, a unified directory entry is reserved under an internet domain (default path: /.well-known/agent-descriptions), and accessing this URL returns a JSON-LD manifest containing URLs of all agent AD documents on that domain. The manifest document type is CollectionPage, which lists the link addresses of each agent description document and can be paginated to support massive numbers of agents. With the active discovery mechanism, any agent only needs to know another's domain name to obtain its agent directory, thereby traversing the entire agent network level by level.
    \setlength{\itemindent}{1em} \item \textbf{Passive Discovery}: Similar to how search engines index web pages, agents can proactively submit their information to specialized search service agents. The specific approach involves search services providing a registration API interface (publicly documented in their agent description), which other agents can call to register their AD document URLs. After receiving a registration request, the search service periodically crawls the URL to obtain the latest agent description and indexes it in its database. When an agent or user queries related services, the search service can return information about registered agents. Through passive discovery, new agents can announce their existence, get indexed by indexing services, and thus be found by more nodes.
\end{itemize}

These two mechanisms complement each other: active discovery focuses on using distributed domain directories to publish agent lists, while passive discovery introduces centralized search agents for index aggregation. All discovery information is represented in JSON-LD format, consistent with the ADP specification. Regardless of the acquisition path, the discovery results ultimately point to the description documents of various agents.

Through the Agent Discovery Protocol, ANP ensures the openness of the agent network: newly joined agents won't silently become information islands. As long as they follow the protocol to publish or register their description documents, they can be retrieved in the network. This provides foundational support for the formation of a large-scale agent ecosystem.

Notably, the Agent Discovery Protocol also specifies standard methods for result pagination and linking to efficiently manage massive agent information. When a domain has many agents, the .well-known/agent-descriptions document can link to subsequent pages through the next field, enabling incremental loading. Search services can adopt similar pagination mechanisms when returning query results. These details ensure the scalability and robustness of the discovery process.

Through the combination of the Agent Description Protocol and the Discovery Protocol, ANP achieves a closed loop at the application layer: agents use ADP to describe themselves and the Agent Discovery Protocol to expose themselves and discover others. This mechanism addresses the challenge of "how to recognize each other" in the agent internet, meeting the requirements for agent interconnection mentioned earlier. Combined with the secure identity layer and flexible meta-protocol layer, ANP provides a complete solution for establishing open, efficient, and trustworthy connections between agents.

\subsection{AI-Native Data Network}
Existing internet infrastructure is designed for human access. Webpages are connected through hyperlinks (HTML+HTTP), with information presented in the form of pages, forms, images, and other visual interfaces, primarily serving human browsing and operation. However, this human-centered connection mode is not suitable for efficient access and processing by AI agents. For AI to use the existing internet, it needs to simulate human behavior (clicking, browsing, parsing webpages), which greatly limits the capabilities of agents.

\begin{figure}
    \centering
    \includegraphics[width=1.0\linewidth]{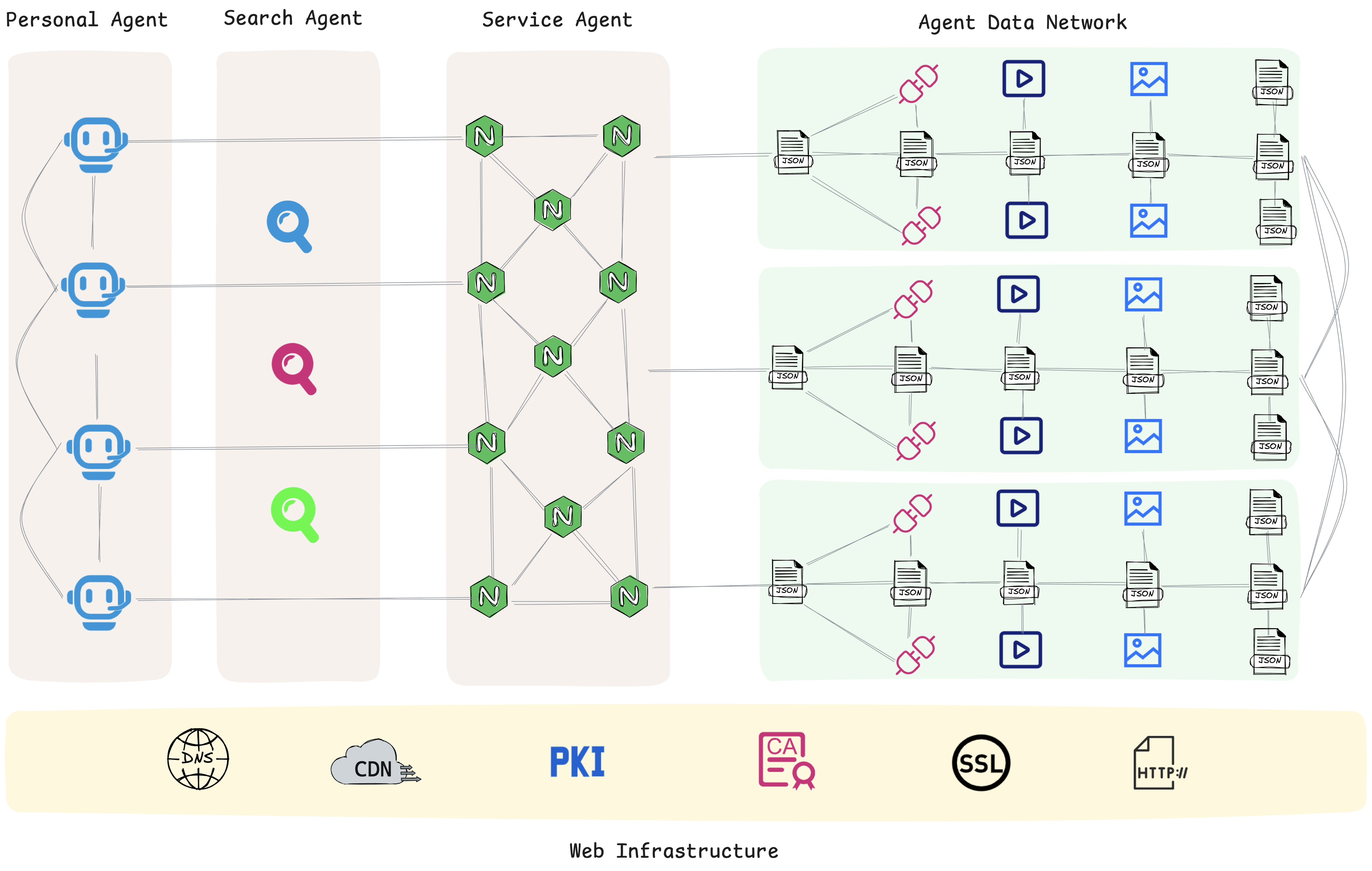}
    \caption{Illustration of the structure of AI-native data network.}
    \label{fig:ai-native-network}
\end{figure}

To address this issue, Agent Network Protocol (ANP) proposes an AI-native data network structure, which is depicted in Figure~\ref{fig:ai-native-network}. Through the Agent Description Protocol (ADP) and Agent Discovery Protocol, each agent can publicize its service interfaces, capability information, and data resources in a structured, standardized way, and automatically expose itself to the entire agent network through open discovery mechanisms.

Under this framework, all agents and their carried data resources (such as text, APIs, audio/video, images, etc.) form an open network that is machine-friendly:
{
\begin{itemize}
    \setlength{\setlength{\itemindent}{1em} \itemindent}{1em} \setlength{\itemindent}{1em} \item Each node is a describable, discoverable, and callable agent or data unit.
    \setlength{\setlength{\itemindent}{1em} \itemindent}{1em} \setlength{\itemindent}{1em} \item Each link is a semantically clear, structurally uniform protocol connection, rather than just hypertext for human reading.
    \setlength{\setlength{\itemindent}{1em} \itemindent}{1em} \setlength{\itemindent}{1em} \item Agents can quickly retrieve, understand, and call resources and services provided by other agents based on unified standards.
\end{itemize}
}

This AI-native data network enables agents to directly, efficiently, and batch-access widely distributed capabilities and knowledge on the internet without relying on webpage crawling or interface simulation, greatly liberating the potential of AI for autonomous learning, reasoning, and collaboration.

As more personal agents, service agents, and search agents connect through ANP, this network with agents as nodes and structured data as connections will continue to grow, eventually forming a global, dynamically evolving agent internet.

%% file: Sections/section4.tex
\section{Security and Privacy}
In the agent internet, security and privacy protection are fundamental and non-negotiable requirements. Agent Network Protocol (ANP) incorporates \textbf{identity authentication security, operation authorization distinction, privacy protection, and fine-grained permission control} as core mechanisms from its initial design, ensuring trusted communication and secure data sharing among agents in an open network environment.

This section systematically explains ANP's security and privacy design from three aspects: identity and authorization management, privacy protection mechanisms, and communication security.

\subsection{Distinction Between Human Authorization and Agent Authorization}

To ensure human control over sensitive operations, ANP introduces a dedicated verification method called \codeinline{humanAuthorization} in DID documents.

\begin{itemize}
    \item \textbf{Low-risk operations} (such as querying public information, browsing data) allow agents to authorize automatically using their autonomous keys without human intervention.
    \item \textbf{High-risk operations} (such as fund transfers, important data submissions, privacy information disclosure) must be explicitly authorized by human users.
\end{itemize}

When executing high-risk requests, user agents must sign using the \codeinline{humanAuthorization} method. This process requires agents to first initiate an authorization request to the human user, and only after explicit confirmation by the user (for example, through biometric verification, password verification, or hardware security module confirmation) can they call the private key bound to that method for signing and submission.

This mechanism effectively ensures that \textbf{the final step of important decision operations must be driven by human will}, preventing agents from being misused or executing high-risk instructions without awareness.

\subsection{Private Key Management and Permission Isolation}
Agent developers need to implement strict permission isolation and security management for various keys, especially the \codeinline{humanAuthorization} key, including but not limited to:

\begin{itemize}
    \item \textbf{Hierarchical management}: Separate ordinary request keys from highly sensitive keys, managing and using them separately.
    \item \textbf{Local encrypted storage}: Private keys should be stored in securely encrypted local devices (such as TEE, HSM) or protected key management systems.
    \item \textbf{Operation logging}: Complete operation logs should be recorded for each sensitive signing operation to facilitate post-event auditing and tracking.
\end{itemize}

These mechanisms maximize protection against risks of key leakage, theft, or misuse.

\subsection{Multi-DID Strategy and Fine-grained Privacy Protection}
To enhance privacy protection and anonymity, ANP recommends that users and agents adopt a \textbf{multi-DID management strategy}. Specific approaches include:

\begin{itemize}
    \item \textbf{Separation of primary DID and sub-DIDs}: The primary DID is used to maintain long-term social relationships (such as friends, business partners), while independent sub-DIDs are generated for different application scenarios (such as e-commerce shopping, food delivery, online services).
    \item \textbf{Local encrypted storage}: Private keys should be stored in securely encrypted local devices (such as TEE, HSM) or protected key management systems.
    \item \textbf{Dynamic verification}: Accessing highly sensitive private keys requires additional dynamic verification (such as fingerprint, facial recognition, one-time passwords).
    \item \textbf{Operation logging}: Complete operation logs should be recorded for each sensitive signing operation to facilitate post-event auditing and tracking.
\end{itemize}

These mechanisms maximize protection against risks of key leakage, theft, or misuse.

\subsection{Minimal Information Disclosure and Communication Privacy}
In communication between agents, ANP emphasizes adherence to the \textbf{Minimal Disclosure Principle}:

\begin{itemize}
    \item Agents should only transmit necessary information fields when completing requests, avoiding unrelated data leakage.
    \item Sensitive fields should be transmitted using end-to-end encryption, ensuring that data content remains protected even if communication links are intercepted.
    \item All communication sessions should be bound to corresponding identity verification information to prevent man-in-the-middle attacks and communication forgery.
\end{itemize}

Additionally, ANP encourages the further adoption of anonymous communication technologies at the application layer (such as verifiable encryption, selective disclosure credentials) to strengthen the protection and access control of private data.

%% file: Sections/section5.tex
\section{Future Prospects: Reshaping the Open Network Through Connection}
The evolution of the internet profoundly validates a core concept: "Connection is Power." In a truly open, interconnected network, free interaction between nodes maximizes innovation potential and creates enormous value. However, today's internet ecosystem is increasingly dominated by a few large platforms, with vast amounts of data and services confined within closed "digital silos," concentrating connection power in the hands of a few tech giants.

The advent of the Agentic Web provides a historic opportunity to reshape this imbalance. Our goal is to drive the internet from its current closed, fragmented state back to its open, freely connected origins. In the future agent network, every agent will simultaneously play the dual roles of information consumer and service provider. More importantly, each node should be able to discover, connect, and interact with any other node in the network without barriers. This vision of universal interconnection will greatly reduce the barriers to information flow and collaboration, returning the power of connection to every user and individual agent.

This marks an important shift: from platform-centered closed ecosystems to protocol-centered open ecosystems. In the latter, value acquisition depends more on the unique capabilities and contributions that participants bring to the network by following open protocols, rather than depending on control over a closed platform. This transformation will stimulate more intense application-layer innovation and competition, as the key to success is no longer "locking in" users but providing superior agent services, similar to the innovation patterns historically promoted by open protocols such as TCP/IP and SMTP.

Building the agent internet network is a grand undertaking that requires extensive collaboration and collective effort. As a foundational protocol framework, ANP's success depends on adoption, implementation, and continuous contribution from the developer community. We invite all researchers, developers, enterprises, and organizations interested in agent technology and the future of the open internet to participate in the development, testing, and application promotion of ANP, working together to build a better future of efficient agent collaboration.